\documentclass{emulateapj}
\usepackage{amssymb}
\usepackage{amsmath,bm}
\usepackage{graphicx}
\usepackage{CJK}
\usepackage[normalem]{ulem}
\usepackage[dvips]{color}
 \usepackage{multirow}
 \usepackage{array}
 \usepackage{dcolumn}% Align table columns on decimal point
\newcommand{\Msun}{$\mathrm{M}_{\odot}$}

\renewcommand\sout{\bgroup \color{red} \ULdepth=-.5ex \ULset}

\begin{document}
\title{Quark matter symmetry energy and quark stars}

\author{Peng-Cheng Chu$^{1}$,  Lie-Wen Chen$^{*1,2}$}

\affil{$^1$ Department of Physics and Astronomy and Shanghai Key Laboratory for Particle
Physics and Cosmology, Shanghai Jiao Tong University, Shanghai 200240, China}
\affil{$^2$ Center of Theoretical Nuclear Physics, National Laboratory of Heavy Ion
Accelerator, Lanzhou 730000, China}
\altaffiltext{*}{Corresponding author, email: lwchen$@$sjtu.edu.cn}

\begin{abstract}
We extend the confined-density-dependent-mass (CDDM) model to include isospin dependence
of the equivalent quark mass. Within the confined-isospin-density-dependent-mass (CIDDM) model,
we study the quark matter symmetry energy, the stability of strange quark matter,
and the properties of quark stars. We find that including isospin dependence
of the equivalent quark mass can significantly influence the quark matter symmetry
energy as well as the properties of strange quark matter and quark stars. While the recently
discovered large mass pulsars PSR J1614-2230 and PSR J0348+0432 with masses around $2M_{\odot}$
cannot be quark stars within the CDDM model, they can be well described by quark stars in the CIDDM
model. In particular, our results indicate that the two-flavor $u$-$d$ quark matter
symmetry energy should be at least about twice that of a free quark gas or normal
quark matter within conventional Nambu-Jona-Lasinio model in order to describe the
PSR J1614-2230 and PSR J0348+0432 as quark stars.
\end{abstract}


\section{Introduction}

One of fundamental issues in contemporary nuclear physics, astrophysics,
and cosmology is to investigate the properties of strong interaction matter,
especially its equation of state (EOS), which plays a central role in
understanding the nuclear structures and reactions, many critical issues in
astrophysics, and the matter state at early universe. Quantum chromodynamics
(QCD) is believed to be the fundamental theory for the strong
interaction. Although the perturbative QCD (pQCD) has achieved impressive success in
describing high energy processes, the direct application of QCD to lower
energy phenomena remains a big challenge in the community due to the
complicated non-perturbative feature of QCD~\citep{Fuk11}. The \emph{ab initio}
lattice QCD (LQCD) numerical Monte Carlo calculations provide a solid basis for
our knowledge of strong interaction matter at finite-temperature regime
with zero baryon density (baryon chemical potential). However, the regime of
finite baryon density is still inaccessible by Monte Carlo because of the
Fermion sign problem~\citep{Bar86}.

In terrestrial laboratory, heavy ion collisions (HIC's) provide a unique
tool to explore the properties of strong interaction matter. The experiments
of high energy HIC's performed (or being performed) in the Relativistic Heavy
Ion Collider (RHIC) at BNL and the Large Hadron Collider (LHC) at CERN have
revealed many interesting features of strong interaction matter at zero
baryon density and high temperature. Instead of the original picture of a
hot ideal gas of non-interacting deconfined quarks and gluons at zero baryon
density and high temperature, the experimental data support a new picture that
quarks and gluons form a strongly interacting system, just like a perfect liquid,
in which non-perturbative physics plays an important role~\citep{Tan09}.
On the other hand, the properties of strong interaction matter at higher
baryon density regions can be explored by the beam-energy scan program at
RHIC which aims to give a detailed picture of QCD phase structure, especially
to locate the so-called QCD critical point~\citep{Ste98}. Knowledge of strong
interaction matter at high baryon density regions can be further complemented
by future experiments planned in the Facility for Antiproton and Ion Research
(FAIR) at GSI and the Nuclotron-based Ion Collider Facility (NICA) at JINR.

In nature, the compact stars provide another way to explore the properties
of strong interaction matter at high baryon density (and low temperature).
Neutron stars (NS's) have been shown to provide natural testing grounds of
our knowledge about the EOS of neutron-rich nuclear matter~\citep{Lat04,Ste05}.
In the interior (or core) of NS's, there may exist hyperons, meson condensations,
and even quark matter. Theoretically, NS's may be converted to (strange) quark
stars (QS's)~\citep{Bom04,Sta07,Her11}, which is made purely of absolutely stable
deconfined $u$, $d$, $s$ quark matter (with some leptons), i.e., strange quark
matter (SQM). Although most of observations related to compact stars can be
explained by the conventional NS models, the QS hypothesis cannot be conclusively
ruled out. One important feature of QS's is that for a fixed mass (especially
for a lighter mass), QS's usually tend to have smaller radii than
NS's ~\citep{Kap01}. It has been argued~\citep{Web05} that the unusual small radii
exacted from observational data support that the compact objects SAX J1808.4每3658, 4U 1728每34, 4U 1820每30,
RX J1856.5每3754 and Her X-1 are QS's rather than NS's. The possible existence
of QS's is one of the most intriguing aspects of modern astrophysics and has
important implications for astrophysics and the strong interaction physics,
especially the properties of SQM which essentially determine the structure of
QS's~\citep{Iva69,Ito70,Bod71,Wit84,Far84,Alc86,Web05}.

The EOS of dense quark matter is usually soft due to the asymptotic freedom of
QCD for quark-quark interactions at extremely high density. In addition, the EOS
of SQM will be further soften due to the addition of $s$ quark which contributes
a new degree of freedom. Therefore, most of quark matter models predict relatively
smaller maximum mass of QS's. Recently, by using the general relativistic Shapiro
delay, the mass of PSR J1614-2230 was precisely measured to be
$1.97\pm0.04M_{\odot}$~\citep{Dem10}. This high mass seems
to rule out conventional QS models (whose EOS's are soft), although some
other models of pulsar-like stars with quark matter can still describe the
large mass pulsar~\citep{Alf03,Bal03,Rus04,Alf05,Alf07,Kla07,Ipp08,Lai11,Wei11,Ave11,Bon12}.
All these models seem to indicate that to obtain a large mass (about $2M_{\odot}$)
pulsar-like star with quark matter, the interaction between quarks should be
very strong, remarkably consistent with the finding in high energy HIC's
that quarks and gluons form a strongly interacting system.

In QS's, the $u$-$d$ quark asymmetry (isospin asymmetry) could be large,
and thus the isovector properties of SQM may play an important role.
Furthermore, the quark matter formed in high energy HIC's at RHIC/LHC
(and future FAIR/NICA) generally also has unequal $u$ and $d$
($\bar u$ and $\bar d$) quark numbers, i.e., it is isospin asymmetric.
In recent years, some interesting features of QCD phase diagram at finite
isospin have been revealed based on LQCD and some phenomenological
models~\citep{Son01,Fra03,Tou03,Kog04,He05a,He05b,DiT06,Zha07,Pag10,Sha12}.
These studies are all related to the isovector properties of quark matter,
which is poorly known, especially at finite baryon density. Therefore,
it is of great interest and critical importance to explore the isovector
properties of quark matter, which is useful for understanding the
properties of QS's, the isospin dependence of hadron-quark phase
transition as well as QCD phase diagram, and the isospin effects of partonic
dynamics in high energy HIC's.

In the present work, we show that QS's provide an excellent astrophysical
laboratory to explore the isovector properties of quark matter at high baryon
density. Through extending the confined-density-dependent-mass (CDDM)
model to include isospin dependence of the equivalent quark mass, we
investigate the quark matter symmetry energy and the properties of SQM and QS's.
We find that, although the maximum mass of QS's within the original CDDM
model is significantly smaller than $2 M_{\odot}$, the isospin dependence of
the equivalent quark mass introduced in the extended
confined-isospin-density-dependent-mass (CIDDM) model can significantly
influence the properties of SQM and the large mass pulsar with mass of $2 M_{\odot}$
can be well described by a QS if appropriate isospin dependence of the equivalent
quark mass is applied.

\section{The theoretical formulism}

\subsection{The confined isospin and density dependent mass model}
\label{Theory}

According to the Bodmer-Witten-Terazawa hypothesis~\citep{Wit84,Web05},
SQM might be the true ground state of QCD matter (i.e., the strong interaction
matter) and is absolutely stable. Furthermore, Farhi and Jaffe found that SQM is
stable near nuclear saturation density for large model parameter space~\citep{Far84}.
The properties of SQM generally cannot be calculated directly from pQCD or LQCD,
because SQM has finite baryon density and its energy scale is not very high.
To understand the properties of SQM, people have built some QCD-inspired
effective phenomenological models, such as the MIT bag model~\citep{Cho74,Far84,Alc86,Alf05,Web05},
the Nambu-Jona-Lasinio (NJL) model~\citep{Reh96,Han01,Rus04,Men06}, the pQCD
approach~\citep{Fre77,Fre78,Fra01,Fra02,Fra05,Fra06,Kur10}, the Dyson-Schwinger
approach~\citep{Rob94,Zon05,Qin11,Li11}, the CDDM
model~\citep{Fow81,Cha89,Cha91,Cha93,Cha96,Ben95,Pen99,Pen00,Pen08,Zha02,Wen05,Mao06,Wu08,Yin08},
and the quasi-particle model~\citep{Sch97,Sch98,Pes00,Hor04,Alf07}. At extremely
high baryon density, SQM could be in color-flavor-locked (CFL) state~\citep{Raj00}
in which the current masses of $u$, $d$ and $s$ quarks become less important
compared with their chemical potentials and the quarks have equal fractions
with the lepton number density being zero according to charge neutrality.

In quark matter models, one of most important things is to treat
quark confinement. The MIT bag model and its density dependent versions provide
a popular way to treat quark confinement. Another popular way to treat quark
confinement is to vary the interaction part of quark mass, such as the
CDDM model and the quasi-particle model. In the present work, we focus on the
CDDM model in which the quark confinement is modeled by the density dependence
of the interaction part of quark mass, i.e., the density dependent equivalent
quark mass.

In the CDDM model, the (equivalent) quark mass in quark matter with baryon density
$n_B$ is usually parameterized as
\begin{equation}
m_q = m_{q0} + m_I = m_{q0} + \frac{D}{{n_B}^z},
\label{qmass}
\end{equation}
where $m_{q0}$ is the quark current mass and $m_I = \frac{D}{{n_B}^z} $ reflects
the quark interactions in quark matter which is assumed to be density dependent,
$z$ is the quark mass scaling parameter, and $D$ is a parameter
determined by stability arguments of SQM. In the original CDDM
model used to study two-flavor $u$-$d$ quark matter~\citep{Fow81}, an inversely
linear quark mass scaling, i.e., $z = 1$ was assumed and the parameter $D$ was
taken to be $3$ times the famous MIT bag constant. The CDDM model was later
extended to include $s$ quarks to investigate the properties of
SQM~\citep{Cha89,Cha91,Cha93,Cha96,Ben95}. Obviously, the CDDM model satisfies two
basic features of QCD, i.e., the asymptotic freedom and quark confinement
through density dependence of the equivalent quark mass, i.e., $\lim_{n_B\to\infty}m_I=0$
and $\lim_{n_B\to0}m_I=\infty$. For two-flavor $u$-$d$ quark matter, the chiral
symmetry is restored at high density due to $\lim_{n_B\to\infty}m_q = 0$ if the
current masses of $u$ and $d$ quarks are neglected.

The density dependence of the interaction part of the quark mass, i.e.,
$m_I = \frac{D}{{n_B}^z}$ is phenomenological in the CDDM model, and in
principle it should be determined by non-perturbative QCD
calculations. Instead of the inversely linear density dependence for
$m_I$ which is based on the bag model argument, a quark mass scaling
parameter of $z = 1/3$ was derived based on the in-medium chiral condensates
and linear confinement~\citep{Pen99} and has been widely used for
exploring the properties of SQM and QS's since then~\citep{Lug03,Zhe04,Pen06,Wen07,Pen08,LiA11}.
In a recent work~\citep{LiA10}, Li A. et al. investigated the stability of SQM and the
properties of the corresponding QS's for a wide range of quark mass scalings.
Their results indicate that the resulting maximum mass of QS's always lies between
$1.5M_{\odot}$ and $1.8M_{\odot}$ for all the scalings chosen there. This
implies that the large mass pulsar PSR J1614-2230 with
a mass of $1.97\pm 0.04M_{\odot}$, cannot be a QS within the CDDM model.
In particular, the maximum mass with scaling parameter $z = 1/3$ is only
about $1.65M_{\odot}$, significantly less than $1.97\pm 0.04M_{\odot}$.

Physically, the quark-quark effective interaction in quark matter should be
isospin dependent. Based on chiral perturbation theory, it has been shown
recently~\citep{Kai09} that the in-medium density dependent
chiral condensates are significantly dependent on the isospin. The isospin
dependence of the in-medium chiral condensates can also be seen from the QCD
sum rules~\citep{Dru04,Jeo12}. In addition, the quark-quark interaction in
quark matter will be screened due to pair creation and infrared divergence
and the (Debye) screening length is also isospin dependent~\citep{Dey98}. These
features imply that the equivalent quark mass in Eq. (\ref{qmass}) should
be isospin dependent which is neglected in the CDDM model. However, the
detailed form of isospin dependence of the equivalent quark mass is unknown,
and in principle it should be determined by non-perturbative QCD calculations.
In the present exploratory work, we extend the CDDM model to include the isospin
dependence of the quark-quark effective interactions by assuming a phenomenological
parametrization form which respects the basic features of QCD, such as the
asymptotic freedom, quark confinement and isospin symmetry, for the equivalent
quark mass in isospin asymmetric quark matter with isospin asymmetry $\delta $, i.e.,
\begin{eqnarray}
m_q &=& m_{q_0} + m_{I} + m_{iso} \notag \\
&=& m_{q_0} + \frac{D}{{n_B}^z} - \tau_q \delta {D_I}n_B^{\alpha}e^{-\beta n_B},
\label{mqiso}
\end{eqnarray}
where $D_I$, $\alpha $, and $\beta $ are parameters determining isospin
dependence of the quark-quark effective interactions in quark matter, $\tau_q $ is the
isospin quantum number of quarks and here we set $\tau_q = 1$ for $q=u$ ($u$ quarks),
$\tau_q = -1$ for $q=d$ ($d$ quarks), and $\tau_q = 0$ for $q=s$ ($s$ quarks).
The isospin asymmetry is defined as
\begin{equation}
\delta = 3\frac{n_d-n_u}{n_d+n_u},
\label{delta}
\end{equation}
which equals to $-n_3/n_B$ with the isospin density $n_3 = n_u-n_d$ and
$n_B = (n_u+n_d)/3$ for two-flavor $u$-$d$ quark matter. The above definition
of $\delta $ for quark matter has been extensively used in the
literature~\citep{DiT06,Pag10,DiT10,Sha12}. We note that one has
$\delta = 1$ ($-1$) for quark matter converted by pure neutron (proton) matter
according to the nucleon constituent quark structure, consistent with the
conventional definition for nuclear matter, i.e., $\frac{\rho_n -\rho_p}{\rho_n +\rho_p}=-n_3/n_B$.

In Eq. (\ref{mqiso}), the last term $m_{iso}$ provides a simple and
convenient phenomenological parametrization form for isospin dependence
of the equivalent quark mass respecting the asymptotic freedom and quark
confinement. Indeed, one can see that the quark confinement condition
$\lim_{n_B\to0}m_q=\infty$ can be guaranteed if $\alpha >0$ or $\alpha =0$.
For $\alpha =0$, in particular, a finite isospin splitting of the equivalent
quark mass will appear even at zero baryon density. In addition, if $\beta >0$,
one then has $\lim_{n_B \to \infty} m_{iso} = 0$ and thus the asymptotic
freedom $\lim_{n_B\to \infty} m_q=m_{q0}$ is satisfied. In general, therefore,
the parameter $\alpha $ should be nonnegative and the parameter $\beta $ should
be positive in Eq.~(\ref{mqiso}). Using different values of $\alpha $ and
$\beta $ can flexibly mimic different density dependences for the isospin dependent
equivalent quark mass, and thus different density dependences for the quark matter
symmetry energy. In this exploratory work, we determine $\alpha $ and $\beta $
by assuming the density dependence of the quark matter symmetry energy has some
well-known empirical forms. The parameter $D_I$ can be used to adjust the strength
of the isospin dependent equivalent quark mass and thus the strength of the quark
matter symmetry energy. In addition, for the quark mass scaling parameter $z$,
in this work, we mainly focus on $z=1/3$ since it can be derived based on the
in-medium chiral condensates and linear confinement~\citep{Pen99}. However, we
also study how our main results change if the quark mass scaling parameter
$z$ can be varied freely. Obviously, in the extended CIDDM model, the equivalent
quark mass in Eq.~(\ref{mqiso}) satisfies the exchange symmetry between $u$ and
$d$ quarks which is required by isospin symmetry of the strong interaction.
Therefore, the phenomenological parametrization form of the isospin dependent
equivalent quark mass in Eq. (\ref{mqiso}) is quite general and respects the
basic features of QCD. Although other functional forms could be used to
describe the isospin dependence of the equivalent quark mass, the exact form of
$m_{iso}$ is not crucial so long as the functional and its parameters are chosen
to be consistent with the asymptotic freedom, quark confinement and isospin
asymmetry as well as some empirical forms of the symmetry energy.

\subsection{The quark matter symmetry energy}

Similarly to the case of nuclear matter~\citep{LCK08}, the EOS of
quark matter consisting of $u$, $d$, and $s$ quarks, defined by its binding
energy per baryon number, can be expanded in isospin asymmetry $\delta $ as
\begin{equation}
E(n_B ,\delta, n_s)=E_{0}(n_B, n_s)+E_{\mathrm{sym}}(n_B, n_s)\delta ^{2}+\mathcal{O}(\delta ^{4}),
\label{EOSANM}
\end{equation}%
where $E_{0}(n_B, n_s)=E(n_B ,\delta =0, n_s)$ is the binding energy per
baryon number in three-flavor $u$-$d$-$s$ quark matter with equal fraction
of $u$ and $d$ quarks; the quark matter symmetry energy $E_{\mathrm{sym}}(n_B, n_s)$ is
expressed as
\begin{eqnarray}
E_{\mathrm{sym}}(n_B, n_s) &=&\left. \frac{1}{2!}\frac{\partial ^{2}E(n_B
,\delta, n_s)}{\partial \delta ^{2}}\right\vert _{\delta =0}.
\label{QMEsym}
\end{eqnarray}%
In Eq. (\ref{EOSANM}), the absence of odd-order terms in $\delta $ is due to
the exchange symmetry between $u$ and $d$ quarks in quark matter when
one neglects the Coulomb interaction among quarks. The higher-order
coefficients in $\delta $ are usually very small and this has been
verified by the calculations with the model parameters in the present work. Neglecting the
contribution from higher-order terms in Eq. (\ref{EOSANM}) leads to the
empirical parabolic law, i.e.,
$E(n_B ,\delta, n_s)\simeq E_{0}(n_B, n_s)+E_{\mathrm{sym}}(n_B, n_s)\delta ^{2}$
for the EOS of isospin asymmetric quark matter and the quark matter symmetry energy
can thus be extracted approximately from the following expression
\begin{eqnarray}
E_{\mathrm{sym}}(n_B, n_s)\simeq \frac{1}{9}[E(n_B ,\delta =3, n_s)\notag \\
-E(n_B,\delta =0, n_s)].
\end{eqnarray}
Based on the model parameters used in the present work, we have checked
that the above expression is a pretty good approximation. However, we have
still used the exact analytical expression of the quark matter symmetry energy
in all calculations in the following.

In quark matter consisting of $u$, $d$, and $s$ quarks, the baryon number
density is given by $n_B=(n_u+n_d+n_s)/3$ and the quark number density
can be expressed as
\begin{equation}
n_i=\frac{g_i}{2\pi^2}\int_0^{\nu_i}k^2dk=\frac{\nu_i^3}{\pi^2},
\end{equation}%
where $g_i=6$ is the degeneracy factor of quarks and $\nu_i$ is the Fermi
momentum of different quarks ($i=u$, $d$, and $s$). Furthermore, the Fermi
momenta of $u$ and $d$ quarks can be expressed, respectively, as
\begin{eqnarray}
{\nu}_u=
{(1-\delta/3)}^{\frac{1}{3}}\nu\notag, \\
{\nu}_d=
{(1+\delta/3)}^{\frac{1}{3}}\nu,
\end{eqnarray}%
where $\nu$ is the quark Fermi momentum of symmetric $u$-$d$ quark matter at quark
number density $n=2n_u=2n_d$. The total energy density of the $u$-$d$-$s$ quark matter
can then be expressed as
\begin{eqnarray}
\epsilon_{uds} &=& \frac{g}{2\pi^2}\int_0^{{(1-\delta/3)}^{\frac{1}{3}}\nu}
\sqrt{k^2+{m_u}^2}k^2dk\notag \\
&+&\frac{g}{2\pi^2}\int_0^{{(1+\delta/3)}^{\frac{1}{3}}\nu}\sqrt{k^2+{m_d}^2}k^2dk\notag \\
&+&\frac{g}{2\pi^2}\int_0^{\nu_s}\sqrt{k^2+{m_s}^2}k^2dk.
\end{eqnarray}%
Using the isospin and density dependent equivalent quark masses as in Eq. (\ref{mqiso}), one
can obtain analytically the quark matter symmetry energy as
\begin{eqnarray}
E_{\mathrm{sym}}(n_B, n_s) &=& \frac{1}{2}\frac{\partial^2\epsilon_{uds}/{n_B}}{\partial \delta^2}\mid_{\delta=0 }\notag \\
&=& \bigg[\frac{\nu^2+18mD_I{n_B}^{\alpha}e^{-\beta n_B}}{18\sqrt{{\nu}^2+m^2}}\notag \\
&+&A+B\bigg]\frac{3n_B-n_s}{3n_B},
\end{eqnarray}
with
\begin{eqnarray}
A&=&\frac{9m^2}{2{\nu}^2\sqrt{{\nu}^2+{m}^2}}(D_I{n_B}^{\alpha}e^{-\beta n_B})^2,\\
B&=&\frac{9}{4{\nu}^3}\bigg[ \nu\sqrt{{\nu}^2+{m}^2}-3m^2ln\bigg(\frac{\nu \sqrt{{\nu}^2+m^2}}{m}\bigg)\bigg]\notag\\
&\times&(D_I{n_B}^{\alpha}e^{-\beta n_B})^2,
\end{eqnarray}
and $m = m_{u0}$ (or $m_{d0}$)$ + \frac{D}{{n_B}^z}$. In the present work, we assume
$m_{u0} = m_{d0} = 5.5$ MeV and $m_{s0} = 80$ MeV. In the CDDM model, the quark matter
symmetry energy is reduced to
\begin{equation}
E_{\mathrm{sym}}(n_B, n_s)=\frac{1}{18}\frac{{\nu}^2}{\sqrt{{\nu}^2+m^2}}\frac{3n_B-n_s}{3n_B}.
\end{equation}%
It should be noted that the quark matter symmetry energy generally depends on the fraction of
$s$ quarks in quark matter since $s$ quarks contribute to the baryon density $n_B$. For two-flavor
$u$-$d$ quark matter, the quark matter symmetry energy is reduced to the well-known expression,
i.e., $E_{\mathrm{sym}}(n_B )=\frac{1}{18}\frac{{\nu}^2}{\sqrt{{\nu}^2+m^2}}$.

\subsection{Properties of strange quark matter}

For SQM, we assume it is neutrino-free and composed of $u$, $d$, $s$ quarks
and $e^-$ in beta-equilibrium with electric charge neutrality. The weak
beta-equilibrium condition can then be expressed as
\begin{eqnarray}
&&\mu_u+\mu_e=\mu_d=\mu_s,
\end{eqnarray}
where $\mu_i$ ($i=u$, $d$, $s$ and $e^-$) is the chemical potential of the
particles in SQM. Furthermore, the electric charge neutrality condition can
be written as
\begin{eqnarray}
&&\frac{2}{3}n_u=\frac{1}{3}n_d+\frac{1}{3}n_s+n_e.
\end{eqnarray}

The chemical potential of particles in SQM can be obtained as
\begin{eqnarray}
\mu_i=\frac{d{\epsilon}}{d{n_i}}&=&\sqrt{{\nu_i}^2+{m_i}^2}+\sum_j{n_j\frac{\partial{m_j}}{\partial{n_B}}}\frac{\partial{n_B}}{\partial{n_i}}f\left(\frac{\nu_j}{m_j}\right)\notag \\
&+&\sum_j{n_j\frac{\partial{m_j}}{\partial{\delta}}}\frac{\partial{\delta}}{\partial{n_i}}f\left(\frac{\nu_j}{m_j}\right),
\label{mu}
\end{eqnarray}%
with
\begin{equation}
f(x)=\frac{3}{2x^3}\left[x\sqrt{(x^2+1)}+\ln{(x+\sqrt{x^2+1}})\right],
\end{equation}%
and $\epsilon$ is the total energy density of SQM. One can see clearly from
Eq. (\ref{mu}) that the chemical potential of quarks in SQM has two additional
parts compared with the case of free Fermi gas, due to the density and isospin
dependence of the equivalent quark mass, respectively. In particular, the $u$ quark
chemical potential can be expressed analytically as
\begin{align}
\mu_u=&\sqrt{\nu^2+m_u^2} +\frac{1}{3}\sum_{j=u,d,s} n_j f\left(\frac{\nu_j}{m_j}\right)\notag\\
&\times\left[-\frac{zD}{n_B^{(1+z)}}- \tau_j D_I\delta(\alpha n_B^{\alpha-1}-\beta n_B^\alpha)e^{-\beta n_B}\right] \notag\\
&+D_In_B^\alpha e^{-\beta n_B}\left[n_u f\left(\frac{\nu_u}{m_u}\right)-n_d f\left(\frac{\nu_d}{m_d}\right)\right] \notag\\
&\times\frac{6n_d}{(n_u+n_d)^2}.
\end{align}%
For $d$ and $s$ quarks, we have, respectively,
\begin{align}
\mu_d=&\sqrt{\nu^2+m_d^2}+\frac{1}{3}\sum_{j=u,d,s} n_j f\left(\frac{\nu_j}{m_j}\right)\notag\\
&\left[-\frac{zD}{n_B^{(1+z)}}- \tau_j D_I\delta(\alpha n_B^{\alpha-1}-\beta n_B^\alpha)e^{-\beta n_B}\right] \notag\\
&+D_In_B^\alpha e^{-\beta n_B}\left[n_d f\left(\frac{\nu_d}{m_d}\right)-n_u f\left(\frac{\nu_u}{m_u}\right)\right]\notag\\
&\times\frac{6n_u}{(n_u+n_d)^2},
\end{align}%
and
\begin{align}
\mu_s=&\sqrt{{\nu_s}^2+{m_s}^2}+\frac{1}{3}\sum_{j=u,d,s} n_j f\left(\frac{\nu_j}{m_j}\right)\notag\\
&\left[-\frac{zD}{n_B^{(1+z)}}- \tau_j D_I\delta(\alpha n_B^{\alpha-1}-\beta n_B^\alpha)e^{-\beta n_B}\right].
\end{align}%
For electrons, the chemical potential can be expressed as
\begin{align}
\mu_e=&\sqrt{{3\pi^2 \nu_e}^2+{m_e}^2}.
\end{align}%

The pressure of SQM can be given by
\begin{align}
P=&-\epsilon+\sum_{j=u,d,s,e} n_j \mu_j\notag\\
=&-\Omega_0+\sum_{i,j=u,d,s,e}{n_in_j\frac{\partial{m_j}}{\partial{n_B}}}\frac{\partial{n_B}}{\partial{n_i}}f\left(\frac{\nu_j}{m_j}\right)\notag\\
&+\sum_{i,j=u,d,s,e}{n_in_j\frac{\partial{m_j}}{\partial{\delta}}}\frac{\partial{\delta}}{\partial{n_i}}f\left(\frac{\nu_j}{m_j}\right),
\end{align}%
where $-\Omega_0$ is the free-particle contribution and $\Omega_0$ can be expressed analytically as
\begin{align}
\Omega_0 =& - \sum_{j=u,d,s,e}\frac{g_i}{48\pi^2} \bigg[\nu_i\sqrt{{\nu_i}^2+{m_i}^2}(2\nu_i^2-3m_i^2)\notag\\
+& 3m_i^4 \texttt{arcsinh}\left(\frac{\nu_i}{m_i}\right)\bigg].
\end{align}
Because of the additional parts in the quark chemical potentials due to the
density and isospin dependence of the equivalent quark mass, the pressure also has
corresponding additional terms. Including such terms is important for guaranteeing
the thermodynamic self-consistency of the model and the Hugenholtz-Van Hove
theorem is then fulfilled~\citep{Pen99}.

\section{Results and discussions}

\subsection{The quark matter symmetry energy}

\begin{figure}[tbp]
\includegraphics[scale=0.75]{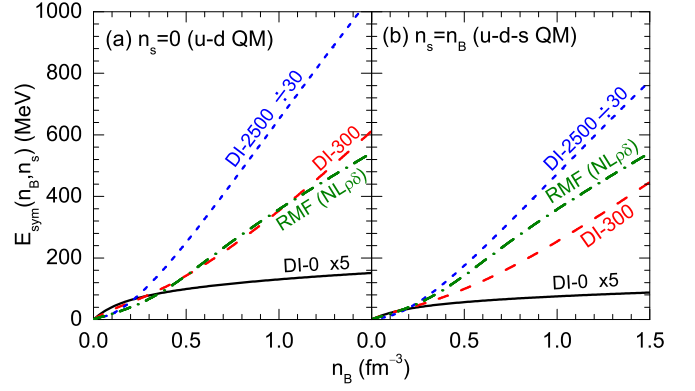}
\caption{(Color online) The quark matter symmetry energy as a function of
baryon number density in the CIDDM model with three parameter sets, i.e.,
DI-0, DI-300, and DI-2500. The two-flavor $u$-$d$ quark matter with $n_s = 0$
(left window) and the $u$-$d$-$s$ quark matter with $n_s = n_B$ (right window)
are considered. The nuclear matter symmetry energy from the RMF model with
interaction NL$\rho\delta$ is also included for comparison. The symmetry energy
values from DI-2500 have been divided by a factor of $30$ while those of DI-0
have been multiplied by a factor of $5$.}
\label{Esym}
\end{figure}

Firstly, we consider the case of $z=1/3$. In Fig.~\ref{Esym}, we show
the baryon number density dependence of the quark
matter symmetry energy in the CIDDM model with three parameter sets,
i.e., DI-0, DI-300, and DI-2500. We have considered two typical cases, i.e.,
two-flavor $u$-$d$ quark matter with $n_s = 0$ and $u$-$d$-$s$ quark matter
with $n_s = n_B$. The latter roughly corresponds to the situation inside QS's
where $s$ quarks may have equal fraction as $u$ and $d$ quarks. For comparison,
we also include in Fig.~\ref{Esym} the nuclear matter symmetry energy from the
covariant relativistic mean field (RMF) model with interaction NL$\rho\delta$~\citep{Liu02}
which includes the isovector-scalar $\delta$ meson field and is obtained by
fitting the empirical properties of asymmetric nuclear matter and describes reasonably
well the binding energies and charge radii of a large number of nuclei~\citep{Gai04}.
Although the density dependence of the nuclear matter symmetry energy is still largely
uncertainty, especially at supersaturation density~\citep{Che12}, the NL$\rho\delta$
result nevertheless represents a typical prediction for the nuclear matter symmetry
energy.

For all the parameter sets DI-0, DI-300 and DI-2500, we have fixed $z=1/3$.
In particular, for the parameter set DI-0, we have $D_I = 0$, $\alpha=0$, $\beta=0$,
and $D=123.328$ MeV$\cdot$fm$^{-3z}$, which corresponds to a typical parameter set
in the CDDM model and is used here mainly for comparison motivation. For the parameter
set DI-300, we have $D_I = 300$ MeV$\cdot$fm$^{3\alpha}$, $\alpha=1$, $\beta=0.1$ fm$^3$,
and $D=115.549$ MeV$\cdot$fm$^{-3z}$, with the values of $D_I$, $\alpha$, and $\beta$
having been obtained so that the quark matter symmetry energy has roughly the same
strength and density dependence as the nuclear matter symmetry energy predicted by
NL$\rho\delta$ while the value of $D$ obtained to guarantee the stability of SQM.
Comparing the results in DI-300 to those in DI-0, one can see how the properties of
SQM and QS's will be influenced if the quark matter symmetry energy has a similar
strength as that of nuclear matter symmetry energy. For the parameter set DI-2500,
we have $D_I = 2500$ MeV$\cdot$fm$^{3\alpha}$, $\alpha=0.8$, $\beta=0.1$ fm$^3$, and
$D=105.084$ MeV$\cdot$fm$^{-3z}$, and as shown in the following these parameter values
have been obtained through searching for the minimum $D_I$ value so that the maximum
mass of a QS can reach to $1.93$\Msun, to be consistent with recently discovered large
mass pulsar PSR J1614-2230 with a mass of $1.97\pm 0.04$\Msun.

It is seen from Fig.~\ref{Esym} that the three parameter sets DI-0, DI-300, and
DI-2500 give very different predictions for the density dependent quark matter
symmetry energy, and thus the three parameter sets allow us to explore the quark
matter symmetry energy effects. In particular, one can see that the symmetry energy
of two-flavor $u$-$d$ quark matter predicted by DI-300 is nicely in agreement with
that of nuclear matter with NL$\rho\delta$ while the $u$-$d$ quark matter symmetry
energy predicted by DI-2500 are about $50$ times the nuclear matter symmetry energy.
On the other hand, the amplitude of the quark matter symmetry energy predicted by
DI-0 is much smaller than that of the nuclear matter symmetry energy. As we will
show later, the parameter set DI-0 predicts roughly the same quark matter symmetry
energy as that of a free quark gas or normal quark matter within conventional NJL
model. In addition, one can see from Fig.~\ref{Esym} that increasing $s$ quark
fraction in $u$-$d$-$s$ quark matter reduces the quark matter symmetry energy as
expected since $s$ quarks contribute to the baryon density $n_B$ while the symmetry
energy is defined by per baryon number.

In the CIDDM model, we can generally increase the amplitude of quark matter symmetry
energy by increasing the $D_I$ value. It should be mentioned that when the value of
$D_I$ parameter is varied, the other three parameters $\alpha, \beta$, and
$D$ (the parameter $D$ if $\alpha$ and $\beta$ have been fixed) usually need
correspondingly readjustment to guarantee the stability of SQM. As we will see in
the following, the three parameter sets DI-0, DI-300, and DI-2500 all satisfy the
stability conditions of SQM, and thus they can be used to study the quark matter
symmetry energy effects on the properties of SQM and QS's.

\subsection{The stability of SQM}

Following Farhi and Jaffe~\citep{Far84}, the absolute stability of SQM requires that the
minimum energy per baryon of SQM should be less than the minimum energy
per baryon of observed stable nuclei, i.e., M($^{56}$Fe)c$^2/56 = 930$ MeV,
and at the same time the minimum energy per baryon of the beta-equilibrium
two-flavor $u$-$d$ quark matter should be larger than $930$ MeV to be
consistent with the standard nuclear physics. These stability conditions
usually put very strong constraints on the value of the parameters in
quark matter models.

\begin{figure}[tbp]
\includegraphics[scale=0.85]{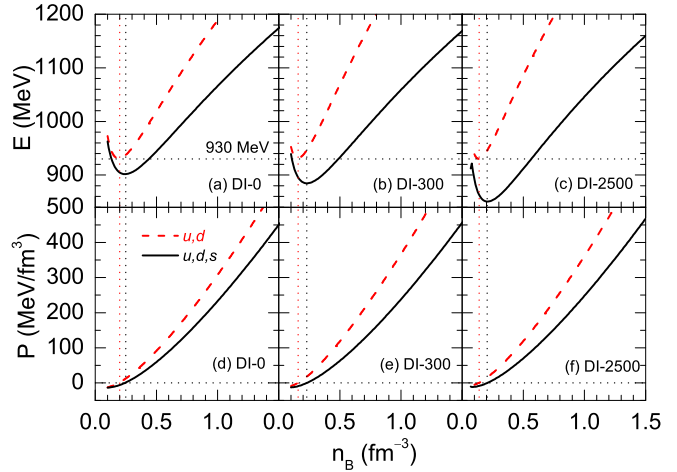}
\caption{(Color online) Energy per baryon and the corresponding pressure
as functions of the baryon density for SQM and two-flavor $u$-$d$
quark matter in $\beta$-equilibrium within the CIDDM model with DI-0, DI-300, and DI-2500.}
\label{EOS}
\end{figure}
Figure~\ref{EOS} shows the energy per baryon and the corresponding pressure
as functions of the baryon density for SQM and two-flavor $u$-$d$ quark matter
in $\beta$-equilibrium within the CIDDM model with DI-0, DI-300, and DI-2500.
One can see that for all the three parameter sets DI-0, DI-300, and DI-2500,
the minimum energy per baryon of the beta-equilibrium two-flavor $u$-$d$ quark
matter is larger than $930$ MeV while that of SQM is less than $930$ MeV,
satisfying the absolute stable conditions of SQM. Furthermore, it is seen from
Fig.~\ref{EOS} that in all cases, the baryon density at the minimum energy per
baryon is exactly the zero-pressure density, which is consistent with the
requirement of thermodynamical self-consistency. In particular, we note that
the zero-pressure density of SQM is $0.24$ fm$^{-3}$, $0.23$ fm$^{-3}$, and
$0.21$ fm$^{-3}$ for DI-0, DI-300, and DI-2500, respectively, which are not
so far from the nuclear matter normal density of about $0.16$ fm$^{-3}$.
Moreover, one can see from Fig.~\ref{EOS} that the stiffness of SQM increases
with the $D_I$ parameter (i.e., the quark matter symmetry energy). In addition,
we have checked the sound speed in the quark matter based on the calculated
pressure and energy density with DI-0, DI-300, and DI-2500, and we find
that the sound speed in all cases is less than the speed of light in vacuum,
and thus satisfying the casuality condition. We would like to note here
that the causality condition is also satisfied for all other EOS's used in
this work for the calculations of QS's.

\begin{figure}[tbp]
\includegraphics[scale=0.85]{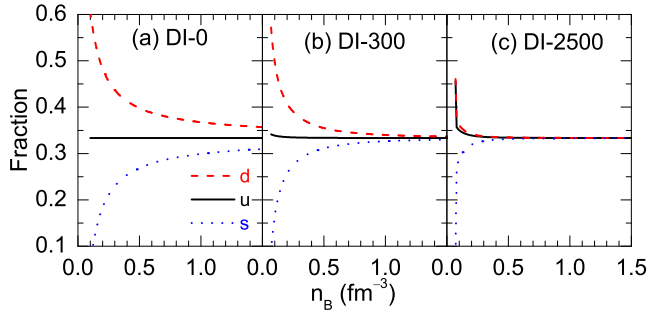}
\caption{(Color online) Quark fraction as a function of the baryon
density in SQM within the CIDDM model
with DI-0, DI-300, and DI-2500}
\label{Frac}
\end{figure}
In Fig.~\ref{Frac}, we show the quark fraction as a function of the baryon
density in SQM within the CIDDM model with DI-0, DI-300, and DI-2500. It is
interesting to see that the difference among $u$, $d$, and $s$ quark fractions
becomes smaller when the quark matter symmetry energy is increased (i.e., from
DI-0 to DI-300, and then to DI-2500). When the quark matter symmetry energy is
not so large (i.e., in the cases of DI-0 and DI-300), the $u$, $d$, and $s$ quark
fractions are significantly different, especially at lower baryon densities,
which leads to a larger isospin asymmetry in SQM. On the other hand, a large
quark matter symmetry energy (i.e., DI-2500) significantly reduces the difference
among $u$, $d$, and $s$ quark fractions. In particular, for DI-2500, it is
remarkable to see that the $u$, $d$, and $s$ quark fractions become essentially
equal and approach a value of about $0.33$ for $n_B \gtrsim 0.4$ fm$^{-3}$,
similar to the results from the picture of CFL state. In neutron star matter,
the similar symmetry energy effect has also been observed, i.e., a larger nuclear
matter symmetry energy will give a larger proton fraction and thus reduce the
difference between neutron and proton fractions in the beta-equilibrium neutron
star matter~\citep{Xu09}.

\begin{figure}[tbp]
\includegraphics[scale=0.85]{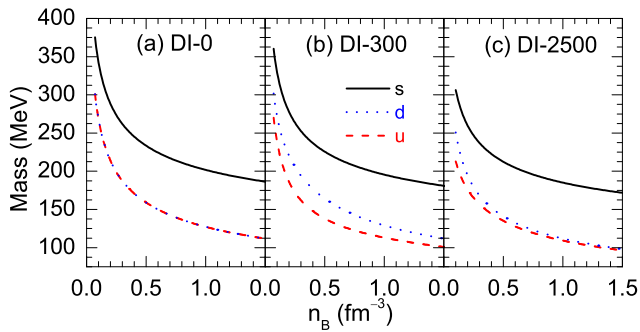}
\caption{(Color online) Equivalent quark mass as a function of the baryon
density in SQM within the CIDDM model
with DI-0, DI-300, and DI-2500.}
\label{Mass}
\end{figure}
Figure~\ref{Mass} shows the equivalent quark mass as a function of the baryon
density in SQM within the CIDDM model with DI-0, DI-300, and DI-2500. It is
seen that in all cases, the equivalent quark mass increases drastically with
decreasing baryon density, reflecting the feature of quark confinement.
Furthermore, interestingly one can see a clear isospin splitting of the $u$
and $d$ equivalent quark masses in SQM for the parameter sets DI-300 and DI-2500,
with $d$ quarks having larger equivalent mass than $u$ quarks. These features
reflect isospin dependence of quark-quark effective interactions in isospin
asymmetric quark matter in the present CIDDM model. It should be mentioned
that the isospin splitting of $u$ and $d$ equivalent quark masses in SQM
with DI-2500 is smaller than that with DI-300 although the former has much
larger symmetry energy (and $D_I$ value) than the latter. This is due to
the fact that the isospin splitting of $u$ and $d$ equivalent quark masses
in SQM also depends on the isospin asymmetry which is much smaller for DI-2500
than for DI-300 as seen in Fig.~\ref{Frac}. In particular, one can see from
Fig.~\ref{Mass} that the isospin splitting of $u$ and $d$ equivalent quark
masses in SQM becomes extremely small at higher baryon densities for DI-2500
due to the extremely small isospin asymmetry in SQM as shown in Fig.~\ref{Frac}.

\subsection{Quark stars}

\begin{figure}[tbp]
\includegraphics[scale=0.88]{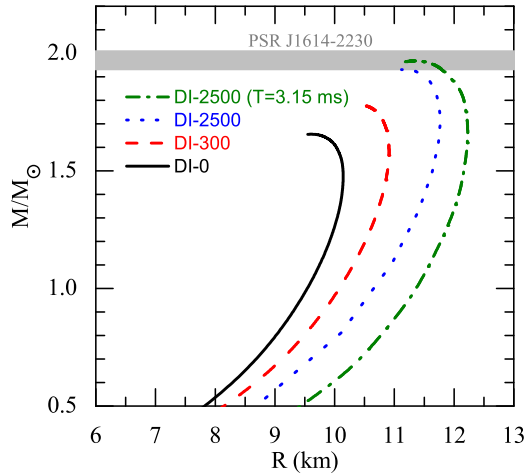}
\caption{(Color online) Mass-radius relation for static quark stars within
the CIDDM model with DI-0, DI-300, and
DI-2500. The result for rotating quark starts with a spin period of 3.15 ms
is also shown for the case of DI-2500 with the radius at the equator. The
shaded band represents the mass of pulsars of $1.97\pm 0.04M_{\odot }$ 
from PSR J1614-2230 \citep{Dem10}.}
\label{MR}
\end{figure}

Using the EOS's of SQM as shown in Fig.~\ref{EOS}, one can obtain the
mass-radius relation of static QS's by solving the Tolman-Oppenheimer-Volkov
equation. Shown in Fig.~\ref{MR} is the mass-radius relation for static
QS's within the CIDDM model with DI-0, DI-300, and DI-2500. Indicated by
the shaded band in Fig.~\ref{MR} is the measured pulsar mass of
$1.97\pm 0.04M_{\odot }$ from PSR J1614-2230~\citep{Dem10}.
For the parameter set DI-2500, we also include in Fig.~\ref{MR} the result for
rotating QS's with a spin period of $3.15$ ms (i.e., the measured value for PSR
J1614-2230), by using the RNS code~\citep{Coo94,Ste95,Kom89} developed by Stergioulas
(available as a public domain program at http://www.gravity.phys.uwm.edu/rns/).
The radius value of the rotating QS's is taken at the equator.

From Fig.~\ref{MR}, one can see for the parameter set DI-0 which
corresponds to the case of the CDDM model and has very small quark matter
symmetry energy, the maximum mass of the static QS's is $1.65M_{\odot }$ and the
corresponding radius is $9.60$ km. Therefore, the maximum mass with DI-0
is significantly smaller than the observed value of $1.97\pm 0.04M_{\odot }$
for PSR J1614-2230. For the parameter set DI-300 whose prediction of quark
matter symmetry energy has almost same amplitude as the nuclear matter symmetry
energy, the maximum mass of static QS's is increased to $1.78M_{\odot }$
with the corresponding radius of $10.40$ km, indicating that increasing the
$D_I$ value (and thus the quark matter symmetry energy) in the CIDDM model
can significantly enhance the maximum mass of static QS's.

However, the maximum mass of QS's with DI-300 is still significantly
smaller than the observed value of $1.97\pm 0.04M_{\odot }$ for PSR J1614-2230.
By further increasing the $D_I$ value and adjusting the
parameters $D$, $\alpha$ and $\beta$ to satisfy the stability condition of SQM
and to obtain a similar density dependence for the quark matter symmetry energy
and the nuclear matter symmetry energy predicted by NL$\rho\delta$,
we find $D_I$ should be larger than $2500$ MeV$\cdot$fm$^{3\alpha}$ so as to
obtain a QS with mass of $1.93M_{\odot }$, and this leads to the parameter
set DI-2500. As shown in Fig.~\ref{MR}, the DI-2500 parameter set gives rise
to a maximum mass of $1.93M_{\odot }$ for static QS's with the corresponding
radius of $11.12$ km, which is consistent with the observed mass of
$1.97\pm 0.04M_{\odot }$ for PSR J1614-2230 within the error bar.
Furthermore, considering the rotation of QS's with a spin period of $3.15$
ms as the measured value for PSR J1614-2230, we obtain the maximum mass of
$1.97M_{\odot }$ for rotating QS's with the corresponding radius of $11.33$
km for the parameter set DI-2500, which is exactly the center value of the
observed mass of $1.97\pm 0.04M_{\odot }$ for PSR J1614-2230. Our results imply that
in the CIDDM model (with a fixed quark mass scaling parameter $z=1/3$),
a larger $D_I$ parameter (and thus a strong isospin dependence of the equivalent
quark mass) is necessary to describe the observed large mass of
$1.97\pm 0.04M_{\odot }$ for a QS, which means the amplitude of the quark matter
symmetry energy should be much larger than that of the nuclear matter symmetry
energy. These features indicate that PSR J1614-2230 could be a QS in the
CIDDM model, and if PSR J1614-2230 were indeed a QS, it can put important
constraint on the isovector properties of quark matter, especially the quark
matter symmetry energy.

\begin{table}
\caption{The maximum mass, the corresponding radius and central baryon
number density of the static quark stars, the maximum rotational frequency
$f_{max}$ for maximum-mass static quark stars as well as the corresponding
gravitational mass and equatorial radius at $f_{max}$, within the CIDDM model with DI-0, DI-300, and DI-2500.}
\begin{tabular}{c|ccc}
        \hline
          & DI-0 & DI-300 & DI-2500 \\
         \hline
        \\$M/M_{\odot} (static)$ & 1.65  & 1.78 & 1.93 \\
        \\$R (km) (static)$ & 9.60 & 10.40 & 11.12 \\
        \\$Central~density (fm^{-3})$ &1.31 & 1.11&1.06 \\
        \\$f_{max}$ (Hz) & 1680 & 1547 & 1458 \\
        \\$M/M_{\odot}~(at~f_{max})$ & 1.78 & 2.12 & 2.43 \\
        \\$R (km)~(equator~at~f_{max})$ & 9.93 & 11.6 & 14.2\\
        \hline
\end{tabular}
\label{Tab1}
\end{table}
For an EOS of SQM, it is physically interesting to determine the maximum mass
of QS's at the maximum rotation frequency $\Omega_{max}$ constrained by the mass shedding and
the secular instability with respect to axisymmetric perturbations, which essentially
corresponds to the maximum mass of rotating QS's that the EOS can support.
The maximum angular frequency $\Omega_{max}$ of a rotating QS can be obtained
from its static mass $M^{stat}_{\odot}$ and radius $R^{stat}_{M\odot}$ by
using the empirical formula proposed by~\citep{Gou99}, i.e.,
$\Omega_{max} = 7730(M^{stat}_{\odot}/M_{\odot})^{1/2}(R^{stat}_{M\odot}/10$km$)^{-3/2}$ rad$\cdot s^{-1}$.

In Table~\ref{Tab1}, we list the maximum rotational frequency $f_{max}$ for
maximum-mass static QS's as well as the corresponding gravitational
mass and equatorial radius at $f_{max}$, within the CIDDM model with
DI-0, DI-300, and DI-2500. For completeness, we also
include in Table~\ref{Tab1} the results of the maximum mass, the
corresponding radius and central baryon number density of the static QS's.
From Table~\ref{Tab1}, one can see the maximum rotational frequency $f_{max}$
decreases with $D_I$ while the corresponding mass and equatorial radius
increase with $D_I$. In particular, for the parameter set DI-2500, we obtain
$f_{max} = 1458$ Hz, and the corresponding mass is $2.43M_{\odot }$ with radius
of $14.2$ km, which essentially corresponds to the maximum mass configuration of
rotating QS's that the parameter set DI-2500 can support.

\subsection{Effects of the quark mass scaling parameter}
\label{scalingZ}

As mentioned before, the quark mass scaling parameter $z$ is phenomenological
in the CDDM model, and in principle it should be determined by non-perturbative
QCD calculations. In the original CDDM model~\citep{Fow81}, an inversely linear
quark mass scaling of $z = 1$ was assumed based on the bag model argument while
a quark mass scaling parameter of $z = 1/3$ was derived based on the
in-medium chiral condensates and linear confinement~\citep{Pen99} and $z = 1/3$
has been used in above calculations. As pointed out by Li A. et al.~\citep{LiA10},
however, the derivation in~\citep{Pen99} is still not well justified since only
the first-order approximation of in-medium chiral condensates was considered
and higher orders of the approximation could non-trivially complicate the quark
mass scaling parameter. Actually, there are also some other quark mass scalings in the
literature~\citep{Dey98,Wan00,Zha02,LiA10}. Therefore, it is interesting to
see how the above calculated results will change if the quark
mass scaling parameter $z$ can be varied freely. As pointed out before, the
CDDM model (i.e., the CIDDM model with $D_I=0$) cannot describe the PSR J1614-2230
as a quark star even though the $z$ parameter can be varied freely~\citep{LiA10}.
In the following, we look for the minimum value of $D_I$ (and thus the smallest
quark matter symmetry energy) that is necessary to support a QS with mass of
$1.93M_{\odot }$ in the CIDDM model.

\begin{figure}[tbp]
\includegraphics[scale=0.76]{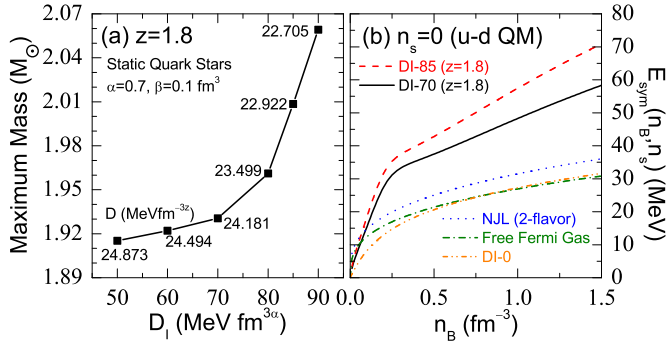}
\caption{(Color online) Left panel: $D_I$ dependence of the maximum mass of static QS's
in the CIDDM model with $z=1.8$. The value of the $D$ parameter at different $D_I$
is obtained so that the QS maximum mass becomes largest. Right panel:
The symmetry energy of two-flavor $u$-$d$ quark matter as a function of baryon
number density in the CIDDM model with DI-70 ($z=1.8$) and DI-85 ($z=1.8$). The results
of DI-0 as well as the symmetry energy of a free quark gas and normal quark matter within
conventional NJL model are also included for comparison.}
\label{EsymMRz}
\end{figure}
To check the effects of the quark mass scaling parameter $z$ and search for the
smallest quark matter symmetry energy to support a QS with mass of $1.93M_{\odot }$,
we assume the quark matter symmetry energy has similar density dependence as that
predicted by the conventional NJL model or that of a free quark gas at higher
baryon densities (here, we consider the density region from $0.25$ fm$^{-3}$ to
$1.5$ fm$^{-3}$) since the lower density EOS will not affect the results of QS's,
and thus we obtain $\alpha=0.7$ and $\beta=0.1$ fm$^3$. Furthermore, for fixed values
of the parameters $D$ and $D_I$, varying the scaling parameter $z$ can significantly
change the maximum mass of QS's and we find that $z=1.8$ generally gives rise
to largest QS maximum mass. For fixed parameters of $\alpha=0.7$, $\beta=0.1$ fm$^3$
and $z=1.8$, and we then search for the minimum value of $D_I$ that can support
a QS with mass of $1.93M_{\odot }$ by varying $D_I$ and $D$. Shown in the left
panel of Fig~\ref{EsymMRz} is the $D_I$ dependence of the maximum mass of
static QS's. The value of the $D$ parameter at different $D_I$ shown in
Fig~\ref{EsymMRz} corresponds to the value at which the QS maximum mass
becomes largest. It is seen from the left panel of Fig~\ref{EsymMRz} that
the maximum mass of QS's is sensitive to the $D_I$ parameter and it increases
with $D_I$. To obtain a QS with mass larger than $1.93M_{\odot }$, we find
the minimum value of $D_I$ should be $70$ MeV$\cdot$fm$^{3\alpha}$, and the
corresponding parameter set is denoted as DI-70 ($z=1.8$). For DI-70 ($z=1.8$),
we thus have $D_I = 70$ MeV$\cdot$fm$^{3\alpha}$, $\alpha=0.7$, $\beta=0.1$ fm$^3$,
$D=24.181$ MeV$\cdot$fm$^{-3z}$, and $z=1.8$. The corresponding radius of
the maximum mass configuration of the QS with DI-70 ($z=1.8$) is $9.69$ km and the
central baryon number density is $1.3$ fm$^{-3}$ while the surface
(zero-pressure point) baryon number density is $0.48$ fm$^{-3}$.

Shown in the right panel of Fig~\ref{EsymMRz} is the density dependence of the
two-flavor $u$-$d$ quark matter symmetry energy in the CIDDM model with DI-70 ($z=1.8$).
For comparison, we also include the results of DI-0 as well as the symmetry energy of a free
quark gas (with current mass of $5.5$ MeV) and normal quark matter within
conventional NJL model~\citep{Reh96}. One can see that the DI-0 parameter set and
the NJL model predict a very similar quark matter symmetry energy as that of the
free quark gas, while the DI-70 ($z=1.8$) parameter set predicts a two times larger quark
matter symmetry energy than the free quark gas but is still significantly smaller
than the nuclear matter symmetry energy predicted by NL$\rho\delta$. Therefore,
our results indicate that, if the $z$ parameter can be varied freely in the CIDDM
model, the quark matter symmetry energy could be smaller than the nuclear matter
symmetry energy but its strength should be at least about twice that of a free quark
gas or normal quark matter within conventional NJL model in order to describe the
PSR J1614-2230 as a quark star. It is interesting to mention that the symmetry
energy of the two-flavor color superconductivity (2SC) phase has been shown to be
about three times that of the normal quark matter phase~\citep{Pag10}, and thus is
close to the symmetry energy predicted by DI-70 ($z=1.8$).

\subsection{The maximum mass of quark stars}

Very recently, a new pulsar PSR J0348+0432 with a mass of $2.01\pm0.04M_{\odot}$
was discovered~\citep{Ant13}. This pulsar is only the second pulsar with a precisely
determined mass around $2M_{\odot}$ after PSR J1614-2230~\citep{Dem10} and gives
the new record of the maximum mass of pulsars. It is thus interesting to examine
if the new pulsar PSR J0348+0432 can be described as a QS within the CIDDM model.
In addition, it is also interesting to see if the CIDDM model can predict even
heavier QS's, which may provide useful implications of future mass measurements
for pulsars.

\begin{figure}[tbp]
\includegraphics[scale=0.88]{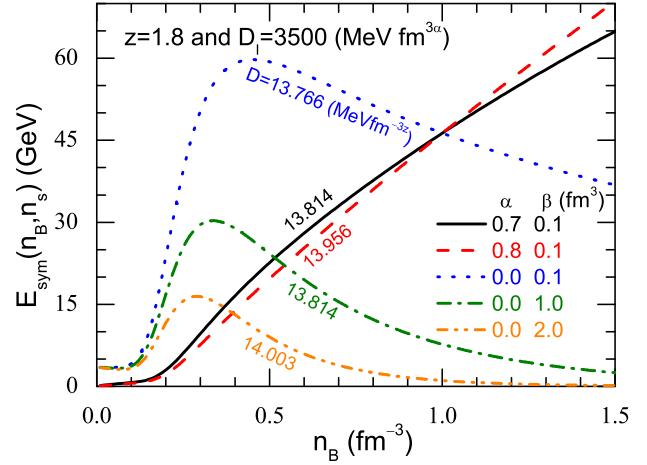}
\caption{(Color online) Density dependence of the two-flavor $u$-$d$
quark matter symmetry energy in the CIDDM model using $z=1.8$ and
$D_I= 3500$ MeV$\cdot$fm$^{3\alpha}$ with different values of $\alpha$
and $\beta$. The value of the $D$ parameter corresponding to the
configuration of the largest maximum mass of static QS's is also
indicted for the different values of $\alpha$ and $\beta$.}
\label{EsymDI3k5}
\end{figure}

As shown above, the maximum mass of static QS's is sensitive to both the quark
matter symmetry energy (via the $D_I$ parameter) and the quark mass scaling
parameter $z$ in the CIDDM model. For $z=1/3$, by increasing the value of the
$D_I$ parameter, we find that the maximum mass of static QS's will saturate at a
value of about $1.96M_{\odot}$ when $D_I$ is larger than about $3000$ MeV$\cdot$fm$^{3\alpha}$
and further increasing the value of the $D_I$ parameter essentially does not change
the maximum mass of static QS's. Furthermore, we note that, when the
value of the $D_I$ parameter is very large (e.g., $D_I= 3000$
MeV$\cdot$fm$^{3\alpha}$), varying the values of $\alpha $ and $\beta $ essentially
has no effects on the maximum mass of static QS's (See also Fig~\ref{EsymDI3k5}
and the related discussions in the following). These interesting features
can be easily understood from the fact that a very large value of $D_I$ gives an
extremely large quark matter symmetry energy which leads to almost equal fraction for
$u$, $d$ and $s$ quarks in the SQM with a very small isospin asymmetry as shown
in Fig.~\ref{Frac}, and consequently the symmetry energy (and the $D_I$ parameter)
effects are strongly suppressed. These results indicate that the CIDDM model with
the $z$ parameter fixed at $1/3$ cannot describe a $2M_{\odot}$ pulsar (e.g., the
pulsar PSR J0348+0432) as a static QS. We note here that the rotation of QS's
with a spin period of $39$ ms as measured for PSR J0348+0432~\citep{Ant13}
essentially has no influences on the maximum mass of QS's.

To further enhance the maximum mass of static QS's, we have to vary the value
of the quark mass scaling parameter $z$ in the CIDDM model. As demonstrated in
Section~\ref{scalingZ}, the quark mass scaling parameter $z$ may significantly
affect the maximum mass of QS's, and a value of $z=1.8$ generally gives the
largest maximum mass of static QS's. As shown in the left panel of
Fig~\ref{EsymMRz}, for $z=1.8$ with $\alpha=0.7$ and $\beta=0.1$ fm$^3$ which
leads to a similar density dependence of the quark matter symmetry energy as
that predicted by the conventional NJL model or that of a free quark gas, one can
find that the minimum value of $D_I$ should be $85$ MeV$\cdot$fm$^{3\alpha}$ to
obtain a QS with mass of $2.01M_{\odot }$, corresponding to the measured center
value for the pulsar PSR J0348+0432, and the corresponding parameter set is
denoted as DI-85 ($z=1.8$). For DI-85 ($z=1.8$), we thus have $D_I = 85$
MeV$\cdot$fm$^{3\alpha}$, $\alpha=0.7$, $\beta=0.1$ fm$^3$, $D=22.922$
MeV$\cdot$fm$^{-3z}$, and $z=1.8$. The corresponding radius of the maximum
mass ($2.01M_{\odot }$) configuration of the QS with DI-85 ($z=1.8$) is $9.98$ km
and the central baryon number density is $1.25$ fm$^{-3}$ while the surface
(zero-pressure point) baryon number density is $0.465$ fm$^{-3}$. The parameter
set DI-85 ($z=1.8$) thus gives the minimum value of $D_I$ (and thus the smallest
quark matter symmetry energy) that is necessary to support a QS with mass of
$2.01M_{\odot }$ in the CIDDM model. The density dependence of the two-flavor
$u$-$d$ quark matter symmetry energy in the CIDDM model with DI-85 ($z=1.8$) is
also plotted in the right panel of Fig~\ref{EsymMRz}. It is seen that the
DI-85 ($z=1.8$) parameter set predicts a slightly larger quark matter symmetry
energy than the DI-70 ($z=1.8$) parameter set and the resulting strength of the
symmetry energy is still about two times larger than that of the free quark gas or
that predicted by the conventional NJL model. Therefore, our results demonstrate
that the CIDDM model can very well describe the new pulsar PSR J0348+0432 as a QS
so long as the $z$ parameter can be varied freely and appropriate isospin
dependence of the equivalent quark mass (and thus the quark matter symmetry
energy) is used.

\begin{figure}[tbp]
\includegraphics[scale=0.75]{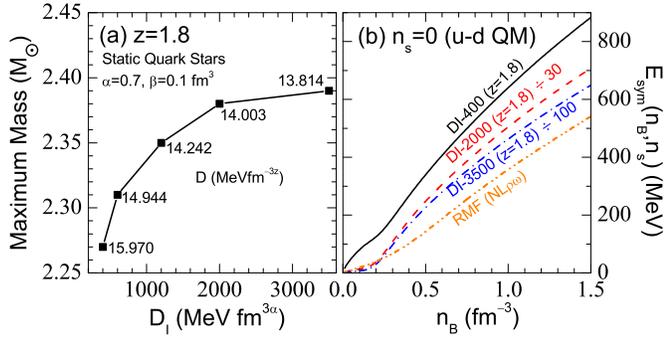}
\caption{(Color online) Same as in Fig~\ref{EsymMRz} but with larger $D_I$ values.
The nuclear matter symmetry energy from the RMF model with interaction NL$\rho\delta$
is also included in the right panel for comparison. The symmetry energy values from
DI-2000 ($z=1.8$) and DI-3500 ($z=1.8$) have been divided by a factor of $30$ and
$100$, respectively.}
\label{MaxMass}
\end{figure}

As shown in the left panel of Fig~\ref{EsymMRz}, the maximum mass of static
QS's increases rapidly with the increment of the $D_I$ parameter. It is thus
interesting to check if there exists a maximum value of the static QS maximum
mass when the $D_I$ parameter further increases. Shown in Fig~\ref{MaxMass} is
the same as in Fig~\ref{EsymMRz} but with larger $D_I$ values. From the left
panel of Fig~\ref{MaxMass}, one can see that the maximum mass of static QS's
still increases rapidly with $D_I$ when $D_I$ is less than about $2000$
MeV$\cdot$fm$^{3\alpha}$. On the other hand, when $D_I$ is larger than
about $2000$ MeV$\cdot$fm$^{3\alpha}$ (which predicts a value of $2.38M_{\odot}$
for the maximum mass of static QS's), the maximum mass of static QS's becomes
insensitive to the $D_I$ parameter, and similarly to the case of $z=1/3$,
the maximum mass of static QS's saturates at a value of about
$2.39M_{\odot}$ when $D_I$ is larger than $3500$ MeV$\cdot$fm$^{3\alpha}$
and further increasing the value of the $D_I$ parameter essentially has no
effects on the maximum mass of static QS's. From the right panel of
Fig~\ref{MaxMass}, one can see that the two-flavor $u$-$d$ quark matter symmetry
energy is much larger than the nuclear matter symmetry energy when $D_I$ is
larger than about $2000$ MeV$\cdot$fm$^{3\alpha}$, and similar results are
also observed in the case of DI-2500 with $z=1/3$ as shown in the left panel
of Fig~\ref{Esym}.

For the results shown in Fig~\ref{MaxMass}, we have fixed $\alpha=0.7$ and
$\beta=0.1$ fm$^3$ to follow the density dependence of the quark matter
symmetry energy of a free quark gas or that predicted by the conventional
NJL model. For $D_I= 3500$ MeV$\cdot$fm$^{3\alpha}$, we find that varying
the values of $\alpha $ and $\beta $ only has very small influence on the
maximum mass of static QS's. In particular, when $\beta$ is fixed at $0.1$
fm$^3$, the maximum mass of static QS's will become $2.39M_{\odot}$ and
$2.40M_{\odot}$ for $\alpha=0.8$ and $\alpha=0$, respectively. Moreover,
when $\alpha$ is fixed at $0$, the maximum mass of static QS's will keep
at the same value of $2.40M_{\odot}$ for both $\beta=1$ and $2$ fm$^3$.
In order to see how different the quark matter symmetry energy becomes with
these various values of $\alpha$ and $\beta$, we plot in Fig~\ref{EsymDI3k5}
the density dependence of the two-flavor $u$-$d$ quark matter symmetry energy
in the CIDDM model using $z=1.8$ and $D_I= 3500$ MeV$\cdot$fm$^{3\alpha}$
with different values of $\alpha$ and $\beta$. Similarly to the left
panel of Fig~\ref{EsymMRz}, the value of the $D$ parameter for different
values of $\alpha$ and $\beta$ is obtained so that the QS maximum mass
becomes largest. As expected for very large $D_I$ values, one can see that,
although the various values of $\alpha$ and $\beta$ indeed give very different
predictions for the two-flavor $u$-$d$ quark matter symmetry energy, all of
them predict almost the same maximum mass of static QS's. We have
also checked the maximum mass of static QS's with the parameter set
DI-85 (z=1.8) by keeping $D_I$ and $D$ fixed but varying $\alpha$ and
$\beta$ as in Fig~\ref{EsymDI3k5}, and our results indicate a similar
tiny effect on the maximum mass of QS's as in the case of using $z=1.8$
and $D_I= 3500$ MeV$\cdot$fm$^{3\alpha}$ (the variation is only about
$0.01M_{\odot}$ ). Due to the finite
isospin splitting of the equivalent quark mass in asymmetric quark matter at
$n_B =0$ for $\alpha = 0$, one can see that the symmetry energy is also finite
at zero baryon number density for $\alpha=0$. From the above results and
discussions, therefore, we conclude that the maximum value of the static QS
maximum mass is about $2.40M_{\odot}$ within the CIDDM model if the $z$
parameter and the strength of the quark matter symmetry energy can be varied
freely.

From the right panel of Fig~\ref{MaxMass}, one can see that an extremely large
two-flavor $u$-$d$ quark matter symmetry energy with its amplitude about $100$
times larger than that of nuclear matter symmetry energy is necessary to describe
a static QS with a mass of $2.40M_{\odot}$ within the CIDDM model with $z=1.8$.
For $z=1/3$, a similar large amplitude of the two-flavor $u$-$d$ quark matter
symmetry energy is also necessary to describe a static QS with a mass of
$1.96M_{\odot}$. Such a large two-flavor $u$-$d$ quark matter symmetry energy
is definitely surprising, and it will be very interesting to investigate its
observable effects in experiments. At this point, we would like to point out that
the absolute stability condition of SQM is still satisfied although the two-flavor
$u$-$d$ quark matter symmetry energy is extremely large, and this usually leads to
almost equal fraction for $u$, $d$ and $s$ quarks in the SQM with a very small
isospin asymmetry, consistent with the picture of CFL state. On the other
hand, the extremely large two-flavor $u$-$d$ quark matter symmetry energy with
very high values of $D_I$ may lead to a negative equivalent quark mass for the
$u$ quark in isospin asymmetric quark matter, and this will indicate a break
down of the model. For example, for DI-2500, the equivalent quark mass of the
$u$ quark will become negative when the baryon density is larger than $0.98$
fm$^{-3}$ if the isospin asymmetry is fixed at $0.05$. The corresponding
critical density will reduce to $0.48$ fm$^{-3}$ if the isospin asymmetry is
fixed at $0.1$. These features imply that the possible existence of quark
matter with high baryon density and large isospin asymmetry may put important
limitations on the amplitude of the two-flavor $u$-$d$ quark matter symmetry
energy or the model parameters in the CIDDM model. In addition, an
extremely large two-flavor $u$-$d$ quark matter symmetry energy may significantly
affect the partonic dynamics in ultra-relativistic heavy ion collisions induced
by neutron-rich nuclei, e.g., Pb + Pb at LHC/CERN or Au + Au at RHIC/BNL, and in
principle the symmetry energy effects in these collisions can be studied
within partonic transport models in which the parton potentials have been
considered~\citep{Son13,XuJ13}. In this case, combining the constraints from
astrophysical observations of heavy QS's and the quark matter symmetry energy
effects in ultra-relativistic heavy ion collisions may provide important
information on both the amplitude of quark matter symmetry energy and the $z$
parameter in the CIDDM model. Our results presented here indicate that $z=1/3$
is ruled out in the CIDDM model if the new pulsar PSR J0348+0432 is a QS.

\section{Conclusion and outlook}
We have extended the confined-density-dependent-mass (CDDM) model in which the
quark confinement is modeled by the density-dependent quark masses to include
isospin dependence of the equivalent quark mass. Within the confined-isospin-density-dependent-mass
(CIDDM) model, we have explored the quark matter symmetry energy, the stability of
strange quark matter, and the properties of quark stars, and found that including
isospin dependence of the equivalent quark mass can significantly change the quark
matter symmetry energy as well as the properties of strange quark matter and quark stars.
We have demonstrated that, although the recently discovered large mass pulsar PSR J1614-2230
with a mass of $1.97\pm0.04M_{\odot}$ cannot be a quark star within the original
isospin-independent CDDM model, it can be well described by a quark star in the CIDDM
model if appropriate isospin dependence of the equivalent quark mass is applied. In
particular, if the density dependent quark mass scaling parameter $z$ is fixed at
$z=1/3$ according to the argument of first-order in-medium chiral condensates and linear
confinement, the equivalent quark mass should be strongly isospin dependent so as to
describe the PSR J1614-2230 as a quark star, leading to that the two-flavor $u$-$d$ quark matter symmetry
energy should be much larger than the nuclear matter symmetry energy. On the other hand,
if the mass scaling parameter $z$ can be varied freely, the two-flavor $u$-$d$ quark matter symmetry energy
could be smaller than the nuclear matter symmetry energy but its strength
should be at least about twice that of a free quark gas or normal quark matter within
conventional Nambu-Jona-Lasinio (NJL) model in order to describe the PSR J1614-2230 as a quark
star. In addition, the most recently discovered large mass pulsar PSR J0348+0432 with a mass of
$2.01\pm0.04M_{\odot}$ can also be described as a quark star within the CIDDM model if the $z$ parameter
can be varied freely and the two-flavor $u$-$d$ quark matter symmetry energy is larger than about twice that of
a free quark gas or normal quark matter within conventional NJL model. Our results have further indicated that $z=1/3$
is ruled out in the CIDDM model if the new pulsar PSR J0348+0432 is a QS.

We have further studied the maximum possible mass of static quark stars within
the CIDDM model and we have found it could be as large as $2.40M_{\odot}$
if the $z$ parameter can be varied freely and the strength of the two-flavor $u$-$d$
quark matter symmetry energy is allowed to be extremely large.

Therefore, our results have demonstrated that the isovector properties of quark
matter may play an important role in understanding the properties of strange quark matter
and quark stars. If PSR J1614-2230 and PSR J0348+0432 were indeed quark stars, they
can put important constraint on the isovector properties of quark matter, especially
the quark matter symmetry energy. In particular, our results have shown that the
two-flavor $u$-$d$ quark matter symmetry energy should be at least about twice that
of a free quark gas or normal quark matter within conventional NJL model in order to
describe the PSR J1614-2230 and PSR J0348+0432 as quark stars.

In the present work, we have mainly focused on the quark matter symmetry energy
and the properties of quark stars within the CIDDM model. In future, it will be
interesting to see how the present results change if other quark matter models
are used, and how the isovector properties of quark matter, especially the quark
matter symmetry energy, will affect other issues such as the quark-hadron phase
transition at finite isospin density, the partonic dynamics in high energy HIC's
induced by neutron-rich nuclei, and so on. These works are in progress.

\section*{Acknowledgments}
The authors would like to thank Wei-Zhou Jiang, Ang Li, Bao-An Li, and
Guang-Xiong Peng for useful discussions. This work was supported in part
by the NNSF of China under Grant Nos. 10975097, 11135011, and 11275125,
the Shanghai Rising-Star Program under grant No. 11QH1401100, the
``Shu Guang" project supported by Shanghai Municipal Education Commission
and Shanghai Education Development Foundation, the Program for Professor
of Special Appointment (Eastern Scholar) at Shanghai Institutions of Higher
Learning, and the Science and Technology Commission of Shanghai
Municipality (11DZ2260700).\\

%%%%%%%%%%%%%%%%%%%%%%%%%%%%%%%%%%%%%%%%%%%%%%%%%%%%%%%%%%%%%%%%%%%

\clearpage
\end{document}